\documentclass[twocolumn,prl,floatfix,showpacs]{revtex4}
\usepackage{color}
\usepackage{epsfig}
\usepackage{graphicx}
\usepackage{amsmath}

\newcommand{\p}{\partial}

\begin{document}

\title{Crack propagation as a free boundary problem}

\author{D. Pilipenko}
\author{R. Spatschek}
\author{E. A. Brener}
\author{H. M\"uller-Krumbhaar}
\affiliation{Institut f\"ur Festk\"orperforschung, Forschungszentrum J\"ulich, D-52425 J\"ulich, Germany}

\pacs{62.20.Mk, 46.15.-x, 46.50.+a, 47.54.-r}

\date{\today}

\begin{abstract}
A newly developed sharp interface model describes crack propagation by a phase transition process.
We solve this free boundary problem numerically and obtain steady state solutions with a self-consistently selected propagation velocity and shape of the crack, provided that elastodynamic effects are taken into account.
Also, we find a saturation of the steady state crack velocity below the Rayleigh speed, tip blunting with increasing driving force and a tip splitting instability above a critical driving force.
\end{abstract}

\maketitle

One of the most challenging riddles in physics and technology is the phenomenon of fracture, as it gives rise to material failure on all scales.
Most fundamentally, the initiation of crack growth is due to a competition of the release of elastic energy and an increase of surface energy, which has been pointed out by Griffith \cite{griffith} and has been used to describe many features of cracks \cite{freund}.
The interpretation of brittle fracture as the successive breaking of atomic bonds is in agreement with models of sharp crack tips, but still the theoretical predictions depend significantly on empirical interaction potentials \cite{hauch}.
On the other hand, if dissipative plastic effects or large scale deformations are important, the crack tips are extended and have a finite tip radius $r_0$.
Here, a detailed description of fracture necessitates equations of motion for each interface point of the extended crack instead of just the mentioned integral energy balance.
A full modeling of fracture should then not only predict the growth velocity but also determine the entire shape of the crack self-consistently.
It was proposed that the characteristic length scale of the crack tip $r_0$ is selected by the threshold of plastic yielding \cite{langer}.
However, approaches of this type require the introduction of dynamic theories of plasticity which are usually much more speculative and less verified than the ordinary linear theory of elasticity.
The lack of suitable equations of motion becomes apparent also in the regime of fast fracture, where the experimentally observed maximum crack speeds are far lower than the theoretically expected Rayleigh speed \cite{freund,sharon}.
Beyond a critical velocity, an unpredictable tip splitting of the crack can occur, producing oscillations of the crack speed \cite{sharon}.

Specific equations of motion have been implemented in various phase field descriptions \cite{aranson,eastgate,karma,karma1,henry,marconi}.
They provide a stimulating approach to describe fracture as a moving boundary problem and go beyond discrete models.
However, additional parameters are introduced in these models in comparison to a conventional linear elastic theory, and the scale of the patterns in the tip region is therefore typically selected by numerical parameters like the phase field interface width.
A purely static elastic description together with macroscopic equations of motion \cite{kassner} provides a well defined sharp interface limit, but suffers from inherent finite time singularities in this limit, which do not allow steady state crack growth.
Based on these insights we recently developed a continuum theory of fracture which resolves this problem by the inclusion of elastodynamic effects \cite{brener,spatschek}.
The main advantage of this model is that it relies only on well established thermodynamical concepts.
Since in the phase field description \cite{spatschek} an extended hierarchy of length scales has to be resolved, expensive large-scale simulations are necessary to predict the steady state growth of cracks:
The system size must be much larger than the crack tip scale, which itself must exceed the phase field interface width and the numerical lattice parameter significantly.
Furthermore, only after a long relaxation time this fully time-dependent description can converge towards a steady state solution.

Here we propose an alternative approach which is specifically dedicated to the fast steady state growth of cracks.
The limit of fully separated length scales is performed analytically, leading to a very efficient numerical scheme.
Furthermore, this approach reaches the valid steady state regime with a self-consistently selected crack shape much faster, which also accelerates the computations dramatically.
The method is based on a multipole expansion technique, and the satisfaction of the elastic boundary conditions on the crack contour reduces to a linear matrix problem, whereas the bulk equations of dynamical elasticity are automatically satisfied.
Nevertheless, finding the correct crack shape and speed remains a difficult nonlinear and nonlocal problem.
In this Letter, we formulate the free boundary problem of crack propagation based on a phase transformation model, and solve it numerically.

{\em Continuum Model of Fracture.}
Imagine that the crack is filled with a soft condensed phase instead of vacuum, and the growth is then interpreted as a first order phase transformation of the hard solid matrix to this soft phase \cite{kassner,brener,spatschek}.
The inner phase becomes stress free if its elastic constants vanish.
For simplicity, we assume the mass density $\rho$ of both phases to be equal;
together with the premise of coherency at the interface this implies that the solid matrix is free of normal and shear stresses at the crack contour, i.e. $\sigma_{nn}=\sigma_{n\tau}=0$, which serves as boundary conditions.
In the bulk, the elastic displacements $u_i$ have to fulfill Newton's equation of motion,
\begin{eqnarray}
\frac{\p \sigma_{ij}}{\p x_j}=\rho \ddot{u}_i.
\label{bulk}
\end{eqnarray}
The difference in the chemical potentials between the two phases at the interface is given by \cite{nozieres}
\begin{equation} \label{steady:eq1}
\Delta\mu=\Omega(\frac{1}{2}\sigma_{ij}\epsilon_{ij}-\gamma\kappa),
\end{equation}
with $\gamma$ being the interfacial energy per unit area;
the interface curvature $\kappa$ is positive if the crack shape is convex;
the atomic volume $\Omega$ appears since the chemical potential is defined as free energy per particle.
For elastically induced phase transitions the motion of the interface is locally expressed by the normal velocity
\begin{equation} \label{steady:eq2}
\upsilon_n=\frac{D}{\Omega\gamma}\Delta\mu
\end{equation}
with a kinetic coefficient D with dimension [D]=m$^2$s$^{-1}$.
We have investigated this minimum model of fracture given by Eqs.~(\ref{bulk})-(\ref{steady:eq2}) by phase field modeling in \cite{spatschek} and have demonstrated that it describes fast crack propagation provided that dynamical elasticity is taken into account.

\begin{figure}
\begin{center}
\epsfig{file=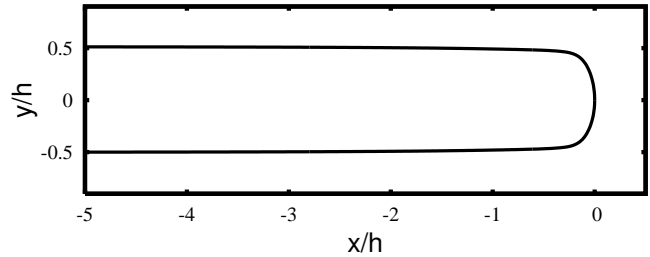, angle=-90, width=9cm}
\caption{Shape of the crack obtained for $\Delta=1.3$; $h$ is the tail opening.}
\label{crack}
\end{center}
\end{figure}

{\em The Multipole Expansion Method.}
We discuss the steady state propagation of a semi-infinite crack in an isotropic medium.
We assume a two-dimensional plane strain situation and mode I loading, which means that the applied tensile forces act perpendicularly to the crack.
The negative $x$ axis is aligned along the semi-infinite crack, see Fig.~\ref{crack}.
We start with the description in the laboratory frame of reference and assume that the crack is moving with a given constant velocity $\upsilon$.
Following Ref.~\cite{freund,rice}, we introduce two real functions $\phi(x,y,t)$ and $\psi(x,y,t)$ which are related to the displacements $u_i$ as follows,
\begin{eqnarray*}
u_x=\frac{\p \phi}{\p x}+\frac{\p \psi}{\p y}, \quad u_y=\frac{\p \phi}{\p y}-\frac{\p \psi}{\p x}.
\end{eqnarray*}
With this decomposition, the bulk equation (\ref{bulk}) decouples to two wave equations,
\begin{eqnarray}
c_d^2\nabla^2\phi=\p^2_{tt}\phi, \quad c_s^2\nabla^2\psi=\p^2_{tt}\psi.
\label{waves}
\end{eqnarray}
Here, $c_d=\sqrt{E(1-\nu)/\rho(1-2\nu)(1+\nu)}$ and $c_s=\sqrt{E/2\rho(1+\nu)}$ are the dilatational and shear sound speed respectively, $E$ is Young's modulus and $\nu$ the Poisson ratio.

In a steady state situation the time derivatives in Eqs.~(\ref{waves}) vanish in a co-moving frame of reference ($x\to x-vt$) and they become Laplace equations there, \begin{equation} \label{laplace}
\frac{\p^2 \phi}{\p x^2}+\frac{\p^2 \phi}{\p y_d^2}=0, \quad \frac{\p^2 \psi}{\p x^2}+\frac{\p^2 \psi}{\p y_s^2}=0.
\end{equation}
We have introduced rescaled coordinates perpendicular to the crack, $y_d=\alpha_d y$ and $y_s=\alpha_s y$, with $\alpha^2_{d}=1-\upsilon^2/c_{d}^2$ and $\alpha^2_{s}=1-\upsilon^2/c_{s}^2$.
For a straight crack with a sharp tip, the solution obeying mode I symmetry and the usual $\sigma\sim r^{-1/2}$ behavior is
\begin{eqnarray*}
\phi=A_0 r_d^{3/2}\cos \frac{3}{2}\theta_d, \quad  \psi=-B_0 r_s^{3/2}\sin \frac{3}{2}\theta_s
\end{eqnarray*}
in rescaled polar coordinates which are related to the co-moving cartesian coordinates via $x=r_d\cos\theta_d=r_s\cos\theta_s$, $y_d=r_d\sin\theta_d$ and $y_s=r_s\sin\theta_s$.
For this mode, the boundary conditions on the straight cut and the matching to the far field behavior demand
\begin{eqnarray}
A_{0}&=&\frac{8(1+\nu)(1+\alpha_s^2)}{\sqrt{2\pi}3E(4\alpha_s\alpha_d-(1+\alpha_s^2)^2)}K_{dyn}, \label{coefA}\\
B_{0}&=& \frac{2\alpha_d}{1+\alpha_s^2}A_{0}, \label{coefB}
\end{eqnarray}
where $K_{dyn}$ is the dynamical mode I stress intensity factor \cite{freund}.

In order to solve the elastodynamic problem of a crack with finite tip radius $r_0$, we use a multipole expansion,
\begin{eqnarray*}
\phi &=& r_d^{3/2} \left[ A_{0}\cos 3\theta_d/2 + \sum_{n=1}^{N=\infty} \frac{A_n}{r_d^{n}} \cos \left( \frac{3}{2}-n \right) \theta_d \right], \\
\psi &=& -r_s^{3/2} \left[ B_{0}\sin 3\theta_s/2 + \sum_{n=1}^{N=\infty} \frac{B_n}{r_s^{n}} \sin \left( \frac{3}{2}-n \right) \theta_s \right].
\end{eqnarray*}
Each eigenmode satisfies the elastodynamic equations (\ref{laplace}).
On macroscopic distances $r$ from the tip, i.e. $r_0 \ll r$, the crack still looks like the semi-infinite mathematical cut and therefore exhibits the same far-field behavior.
Thus, the coefficients $A_{0}$ and $B_{0}$ are determined by Eqs.~(\ref{coefA}) and (\ref{coefB}), whereas all other modes decay fast and do not contribute to the asymptotics.
Consequently, we obtain the formal stress field expansion,
\begin{eqnarray*}
\sigma_{ij}=\frac{K_{dyn}}{(2\pi r)^{1/2}}\left(f_{ij}^{(0)}+\sum_{n=1}^{N=\infty} \frac{A_n f_{ij,d}^{(n)} + B_n f_{ij,s}^{(n)}}{r^n}\right),
\end{eqnarray*}
where $f_{ij,d}^{(n)}(\theta_d, \upsilon)$ and $f_{ij,s}^{(n)}(\theta_s, \upsilon)$ are the universal angular distributions for the dilatational and shear contributions which also depend on the propagation velocity.
The unknown coefficients of expansion can be found by solving the linear problem of fulfilling the boundary conditions $\sigma_{nn}=\sigma_{n\tau}=0$ on the crack contour.
The tangential stress $\sigma_{\tau\tau}$ is determined only through the solution of the elastic problem, and enters into the equations of motion (\ref{steady:eq1}) and (\ref{steady:eq2}), leading to a complicated coupled and nonlocal problem.

The strategy of solution of the problem is as follows:
first, for a given guessed initial crack shape and velocity, we determine the unknown coefficients $A_n$ and $B_n$ from the boundary conditions.
Second, we calculate the chemical potential and the normal velocity at each point of the interface.
Afterwards, the new shape is obtained by advancing the crack according to the local interface velocities.
This procedure is repeated until the shape of the crack in the co-moving frame of reference remains unchanged.
It provides a natural way to solve the problem, as it follows the physical configurations to reach the steady state.
Originally, this idea was developed in the context of dendritic growth \cite{HMKSaito}.
Then the following relation between the local normal velocity and the steady state tip velocity $\upsilon$ holds:
\begin{eqnarray}
\upsilon_n-\upsilon\cos\eta=0,
\label{ss}
\end{eqnarray}
with $\eta$ being the angle between the normal to the crack contour and the $x$ axis.
Alternatively to this ``quasi-dynamical approach'', we also directly solve the nonlinear equation (\ref{ss}) as a functional of the crack shape and the tip velocity $\upsilon$ by Newton's method complemented by Powell's hybrid method \cite{powell,dennis} (``steady state approach'').

Our results are obtained with a finite number of modes $N$, and we performed the extrapolation $N\to\infty$ and found only minor deviations of a few percent from the results for $N=12$ presented here.
An extended discussion of the numerical details will be published elsewhere.

\begin{figure}
\begin{center}
\epsfig{file=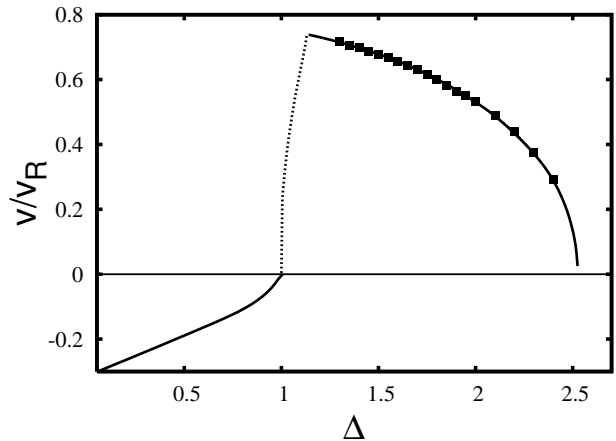, width=6cm,angle=-90}
\caption{Steady state velocity as a function of the dimensionless driving force $\Delta$. The solid line corresponds to the steady state code, the squares to the quasi-dynamical code. Below the point $\Delta_c\approx 1.14$ the dissipation-free solution is selected by a microscopic length scale. Also for $\Delta<1$ the tip scale is not selected, and the presented solution is obtained for the specific parameter $\upsilon_R h/D=10$.}
\label{velocity}
\end{center}
\end{figure}

\begin{figure}
\begin{center}
\epsfig{file=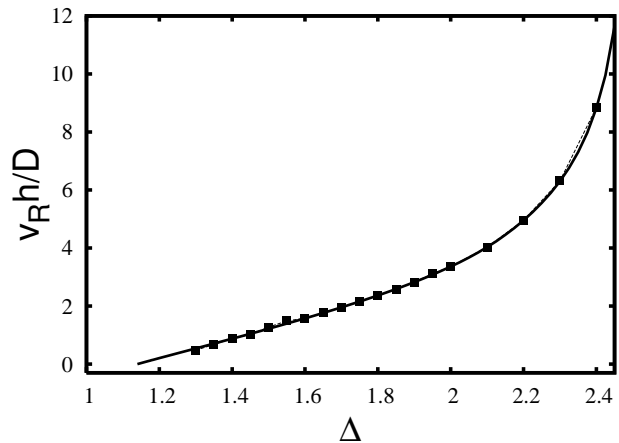, width=6cm,angle=-90}
\caption{The dimensionless tail opening $\upsilon_R h/D$ as a function of the dimensionless driving force $\Delta$. The solid curve corresponds to the steady state code, squares belong to quasi-dynamical calculations.}
\label{diss}
\end{center}
\end{figure}

{\em Results and Discussion.}
We discuss crack growth in a strip geometry, with the width of the strip being very large in comparison to the crack tip scale.
We introduce the dimensionless driving force $\Delta=K_{stat}^2(1-\nu^2)/2E\gamma$, where the static stress intensity factor $K_{stat}$ is assumed to be given and finite.
The relation between the static and the dynamic stress intensity factor can be easily found from energy considerations for the strip geometry in the spirit of \cite{Fineberg}
\begin{equation}
K_{dyn}= K_{stat} \left( (1-\nu) \frac{4\alpha_s\alpha_d-(1+\alpha_s^2)^2}{\alpha_d(1-\alpha_s^2)} \right)^{1/2}.
\end{equation}
The multipole expansion technique confirms that the simple phase transformation model, based on Eqs.~(\ref{bulk})-(\ref{steady:eq2}), provides a selection of the steady state shape and the velocity of the crack.
A typical crack shape is shown in Fig.~\ref{crack} \cite{fn}.
The dimensionless crack velocity $\upsilon/\upsilon_R$ ($\upsilon_R$ is the Rayleigh speed) and the dimensionless crack opening $\upsilon_R h/D$ as function of the driving force $\Delta$ are shown in Figs.~\ref{velocity} and \ref{diss}.
The simulations have been performed with Poisson ratio $\nu=1/3$, which is the only remaining parameter.
All results are obtained both by the ``steady state approach'' and the ``quasi-dynamical code'', and they are in excellent agreement with each other.
Above the Griffith point $\Delta>1$ the shape of the crack obeys the equation $-\upsilon y'=D y''$ in the tail region $x\to -\infty$, because the elastic stresses have decayed there \cite{spatschek}.
Its general solution
\begin{equation} \label{expo}
y(x)=\frac{h}{2}+B\exp(-\upsilon x/D)
\end{equation}
contains a growing exponential which is excluded by the boundary condition of straightness, $y'=0$, and finally leads to the selection of both the steady state propagation velocity and the crack opening $h$ (For more detailed counting arguments, see Refs.~\cite{brener,spatschek}).

In principle, one would expect steady state solutions for crack growth to exist for all driving forces beyond the Griffith point, $\Delta>1$.
However, in the framework of the model, they exist only in the interval $\Delta=1.14-2.5$.
At the limiting value, $\Delta\approx 2.5$, the propagation velocity tends to zero and the length scale $h$ diverges.
Nevertheless, at this point the product $\upsilon h/D$ remains finite, as it is required for finite driving forces.
This termination of the steady state solution is surprising, as one would expect the tip blunting to continue to arbitrarily large values of $\Delta$.

At the lower limit, $\Delta\approx 1.14$, the steady state crack velocity is finite, but the tail opening tends to zero in the framework of the model.
At this point, the dissipation becomes zero, and all energy is converted to surface energy apart from kinetic contributions which are transported through the soft phase and out of the system.
Below this point, i.e.~for $1 < \Delta<1.14$, dissipative solutions do not exist.
Naturally, the tip scale should then be determined by an intrinsic microscopic length scale which is not contained in the present model.
If it was introduced here explicitly, the behavior of the crack speed would behave as depicted by the dotted curve in Fig.~\ref{velocity};
then it would become zero at the Griffith point $\Delta=1$.
Precisely this behavior near the Griffith threshold was observed in phase field simulations \cite{spatschek}, where this cutoff naturally appears as the phase field interface width.

As we have already noted, the length scale of the crack tip becomes large for high driving forces, and therefore at least in this region our macroscopic theory should be valid.
On the other hand, the velocity decays in this regime with increasing driving force, which is an counterintuitive outcome of the model;
nevertheless, the product $\upsilon h/D$ which controls dissipation is monotonically growing.
Physically, it means that the dissipation is mainly increased due to tip blunting instead of a rise of the crack speed.

\begin{figure}
\begin{center}
\epsfig{file=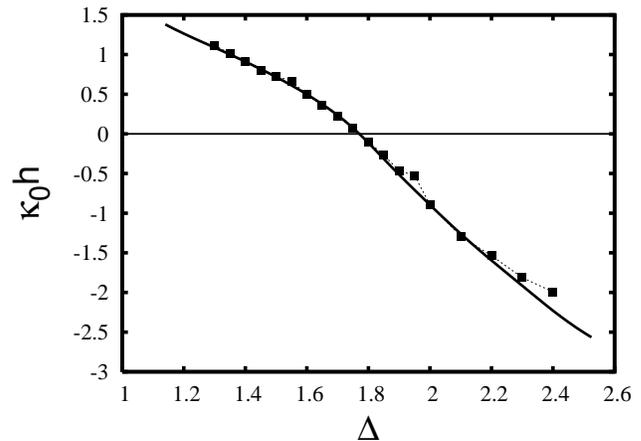, width=6cm,angle=-90}
\caption{Tip curvature $\kappa_0$ as a function of the driving force $\Delta$. The solid curve corresponds to the steady state code, squares belong to quasi-dynamical calculations.}
\label{kappa}
\end{center}
\end{figure}

We suppose that the solutions become unstable against a secondary Asaro-Tiller-Grinfeld instability \cite{ATG} beyond the point $\Delta\approx 1.8$, in agreement with previous conjectures \cite{brener} and phase field simulations \cite{spatschek}.
This is indicated here by the change of sign of the tip curvature $\kappa_0$ as shown in Fig.~\ref{kappa}, corresponding to a tip splitting structure.
Since by construction we confine our investigations to symmetrical steady state solutions, we cannot capture the full asymmetric scenario.
If we released this constraint, the competition between the emerging new tips would lead to complicated non-stationary growth scenarios, as we have seen in phase field simulations \cite{spatschek}.

Finally, we discuss the healing of cracks below the Griffith point, $\Delta<1$;
the velocity $\upsilon$ of the crack becomes negative in this regime.
In contrast to the case of growth, one expects these steady state solutions to exist for arbitrarily prescribed openings $h$ and only the velocity $\upsilon$ to be selected.
This corresponds to the fact that the growing exponential in Eq.~(\ref{expo}) automatically vanishes in the tail region.
This prediction is numerically confirmed by our simulations, see Fig.~\ref{velocity}.
We note that without elastic stresses, i.e. for $\Delta=0$, the problem has a simple analytical solution:
$x(y)/h=(1/\pi)\ln\cos(\pi y/h)$ with velocity $v=\pi D/h$.

Im summary, we have presented a continuum theory of fracture based only on the linear theory of elasticity and a phase transformation process at the crack surface;
we employ a sharp interface method to find steady state solutions of crack growth and are able to predict the growth velocity and the self-consistently selected shape of the crack.
Beyond a critical driving force a negative tip curvature indicates the transition to a tip splitting regime.
The results are in qualitative agreement with phase field simulations \cite{spatschek}.

This work has been supported by the Deutsche Forschungsgemeinschaft under Grant No.~SPP 1120.



\begin{thebibliography}{99}
\bibitem{fn}
We note that the shape is drawn without elastic displacements which should be added to obtain the real shape under load;
for example, the vertical displacement obeys the usual scaling $u_y\sim \sqrt{|x|}$ for large distances from the tip.
\end{thebibliography}
\end{document}